\documentclass[reprint,aip,showpacs,superscriptaddress]{revtex4-1}
\usepackage{graphicx}
\usepackage[usenames]{color}
\usepackage{dcolumn}
\usepackage{bm}
\usepackage{epstopdf}
\makeatletter

\newcommand{\Rmnum}[1]{\expandafter\@slowromancap\romannumeral #1@}
\makeatother

\begin{document}

\title{Influence of electronic structure parameters on the electrical transport and magnetic properties of $Y_{2-x}Bi_xIr_2O_7$ pyrochlore iridates}
\author{Vinod Kumar Dwivedi}
\email{vinodd@iitk.ac.in}
\affiliation{Materials Science Programme, Indian Institute of Technology Kanpur, Kanpur 208016, India}
\author{Soumik Mukhopadhyay}
\affiliation{Department of Physics, Indian Institute of Technology Kanpur, Kanpur 208016, India}
\begin{abstract}
We report the systematic study of structural, magnetic and electrical transport properties of $Y_{2-x}Bi_xIr_2O_7$ (x = 0.0, 0.1, 0.2, 0.3) pyrochlore iridates. The chemical doping enhances electrical conductivity and antiferromagnetic correlation substantially. The replacement of non-magnetic $Y^{3+}$ ion with non-magnetic $Bi^{3+}$ in $Y_2Ir_2O_7$ tends to reduce the octahedral distortion thus enhancing the antiferromagnetic correlation. Raman spectroscopy shows that the $Ir-O$ bond contract slightly and the $R-O'$ bond turn longer as disorder and phononic oscillation are reduced with $Bi$ doping, leading to wider $t_{2g}$ bands, which enhances the electrical conductivity.  Additionally, the enhancement in electrical conductivity and antiferromagnetic correlation with $Bi^{3+}$ doping is attributed to the hybridization between the $Y^{3+}(4p)$/$Bi^{3+}(6s/6p)$ orbital with $Ir^{4+}(5d)$ orbital as a result of enhancement in $Ir-O-Ir$ bond angle and contraction in $Ir-O$ bond length.
\end{abstract}
\maketitle
\section{Introduction}
In recent years, 5d iridates have proved to be a fertile ground for new physics driven by the interplay between the onsite Coulomb repulsion ($U = 0.5-3 eV$), strong crystal field effect ($CF = 1-5 eV$) and spin-orbit coupling ($\lambda_{SOC} = 0.1-1 eV$)~\cite{Krempa1}, potentially leading to novel quantum phases such as a topological Mott insulator~\cite{Pesin}, Weyl semimetal~\cite{Wan1}, axion insulator~\cite{Krempa1,Wan1}, superconductor~\cite{Hanawa}, continuous metal insulator transition (MIT)~\cite{Mandrus} etc. The interplay of competing interactions are controlled~\cite{Koo} \textit{inter-alia} by the $Ir-O-Ir$ bond angle and $Ir-O$ bond length, which can be tuned by chemical~\cite{Harish1,Harish2,Vinod1,Vinod2,Hui1}, physical~\cite{Tafti,Wei,Wei1} or surface~\cite{Vinod3,Vinod4,Vinod5,Vinod6} strain pressure. A small perturbation in $R$-site~\cite{Matsuhira1,Yanagishima,Taira,Zhu,Fukazawa} or $Ir$-site~\cite{Harish1,Harish2,Vinod1,Hui1} in $R_2Ir_2O_7$ (R = Y, Bi or rare earth elements) pyrochlore iridates may easily destabilize the ground state.

Experimental studies show that the $Ir-O-Ir$ bond angle increases as the $R$-site ionic radius increases resulting in a wider $t_{2g}$ bandwidth and MIT~\cite{Koo}. Depending on the substituted element, the SOC and U can be tuned using chemical doping at $Ir$-site. For example, lighter $d$ element compared to $Ir^{4+}(5d^5)$ would reduce the SOC and enhance U simultaneously. For $Y_2Ir_2O_7$, the magnetic properties are determined by the contribution from $Ir^{4+}$ ion. $Y_2Ir_2O_7$ shows a spin-glass like behaviour or canted ferromagnetic transition at temperature T $\sim$ 160K~\cite{Zhu,Fukazawa,Shapiro,Hui2}. Chemical doping of isovalent non-magnetic $Ti^{4+}(3d^0)$ ~\cite{Harish2} and magnetic $Ru^{4+}(4d^4)$ ~\cite{Harish1} ions [both have smaller SOC and larger U than $Ir$] individualy at magnetic $Ir^{4+}(5d^5)$ site in $Y_2Ir_2O_7$ compounds would lead to enhancement in $Ir-O-Ir$ bond angle and reduction in $Ir-O$ bond length. This produces marginal enhancement of electrical conductivity with large increase in antiferromagnetic correlation. Strikingly, chemical doping of magnetic $Cr^{3+}(3d^3)$ ~\cite{Vinod1} ion at magnetic $Ir^{4+}$ site reduces $Ir-O-Ir$ bond angle and enhances $Ir-O$ bond distance subtantially, giving rise to orders of magnitude enhancement of electrical conductivity and ferromagnetism. 

So far as doping at $R$-site is concerned, previous reports on chemical doping of $Ca^{2+}-3p^6$ at $Y^{3+}-4p^6$ site in $Y_2Ir_2O_7$ system alter the electron band width of $Ir-t_{2g}$~\cite{Zhu,Fukazawa} and show enhancement of electrical conductivity along with weakening of antiferromagnetic correlation. On the other hand, introduction of isovalent non-magnetic $Bi^{3+}-6s/6p$ ion at non-magnetic $Y^{3+}$ site in $Y_2Ir_2O_7$ does not alter the $t_{2g}$ band width and yet show MIT~\cite{Aito,Soda}. While studying the magnetization and Hall resistivity below magnetic transition temperature for the $Y_{2-x}Bi_xIr_2O_7$ series with small doping concentration x, Aito et al.~\cite{Aito} and Soda et al.~\cite{Soda} observed hysteretic behaviours, suggesting that the glass-like feature could also be a magnetic state at low temperatures. Additionally, based on $\mu$SR measurements Fernandez et al.~\cite{Fernandez} observed weakening of this behaviour with introduction of Bi content i.e., magnetically ordered state in un-doped compound reduced as Bi doping content increased. It was proposed that in $Y_{2-x}Bi_xIr_2O_7$, the magnetically ordered state associated to $Ir^{4+}$ magnetic ions remained same but the paramagnetic volume fraction got larger and larger. Very recently, a gapped out Weyl-semimetal phase has also been reported in the $Bi$ doped $Y_2Ir_2O_7$ nanoscale system~\cite{Abhishek1,Abhishek2}. These observations underline the fact that there is requirement for deeper understanding of magnetic state as it is possibly connected to the topological Weyl semimetal~\cite{Wan1} state in $Y_2Ir_2O_7$. Despite several reports on chemical doping at non-magnetic $R = Y$ site, detailed analysis of the experimental results as well as finding a correlation between electronic structural parameters and magnetic and electrical transport properties, is presently lacking. Moreover, given the MIT in $Y_{2-x}Bi_xIr_2O_7$ series, the electronic transport behaviour requires to be explored more thoroughly including in the presence of magnetic field.

In the present work, we have studied the interrelation between electronic structural parameters and magnetic and electrical transport properties of $Y_{2-x}Bi_xIr_2O_7$ (x = 0.0, 0.1, 0.2, 0.3) compounds using XRD, FE-SEM, XPS, Raman spectroscopy, dc magnetization and magnetotransport measurements. We show that isovalent doping of non-magnetic $Bi^{3+}$ ion at non-magnetic $Y^{3+}$ site in $Y_2Ir_2O_7$ system alters the electronic structural parameters such as lattice constant, bond angle, bond length and charge states of $Ir$ leading to orders of magnitude enhancement of electrical conductivity and enhanced antiferromagnetic correlations compared to the undoped $Y_2Ir_2O_7$ compound.

\section{Experimental Methods}
The samples $Y_{2-x}Bi_xIr_2O_7$, $x = 0.0, 0.1, 0.2, 0.3$ were synthesized by solid state reaction route following the protocol reported by same authors~\cite{Vinod1,Vinod2,Vinod3,Vinod4,Vinod5,Vinod6,Vinod7,Vinod8} and elsewhere~\cite{Aito,Soda}. High purity stoichiometric mixtures of $IrO_2$ [Alfa-Aesar, Permion (r)], $Bi_2O_3$ [SIGMA-ALDRICH] and $Y_2O_3$ [SIGMA-ALDRICH] with purities of 99.99\% were ground, pressed into pellets and sintered at 1000$^0$C for 24h. The process of grinding, pelletization and sintering (at 1050$^0$C for 24h) were repeated once more for doped compound, i.e. x = 0.1, 0.2, 0.3. For the parent compound i.e. x = 0.0, stoichiometric mixtures of $IrO_2$   and $Y_2O_3$ were ground, pelletized and then sintered between 950$^0$C and 1050$^0$C for 10 days with several intermediate grindings. The quality of compound and crystal structure were analyzed by powder X-ray diffraction (XRD) with Cu K$\alpha$ radiation ($\lambda$= 1.54 $\AA$) at room temperature. The chemical compositions and micro structural morphology of the samples were measured by energy dispersive x-ray spectrometry (EDX) using field emission scanning electron microscope (FE-SEM) [JSM-7100F, JEOL]. The temperature dependence electrical transport properties were measured by conventional four probe technique. The magnetic measurements were carried out using a Quantum design physical property measurement system (PPMS). The X-ray photoelectron spectroscopy (XPS) were measured using a PHI 5000 Versa Probe II scanning XPS system. A micro-Raman measurements were performed using a JY-Horiba micro Raman system having a grating with 1200 grooves/mm (resolution $\sim$ 0.6cm$^{-1}$/pixel). The excitation source was 514.5nm laser line of Ar$^+$ laser with laser power on the sample is 5.86mW. The laser was focused to a spot size of $\sim$ 2-5 $\mu$m on the sample.

\section{Results and Discussion}
\begin{figure}
	\includegraphics[width=8.5cm]{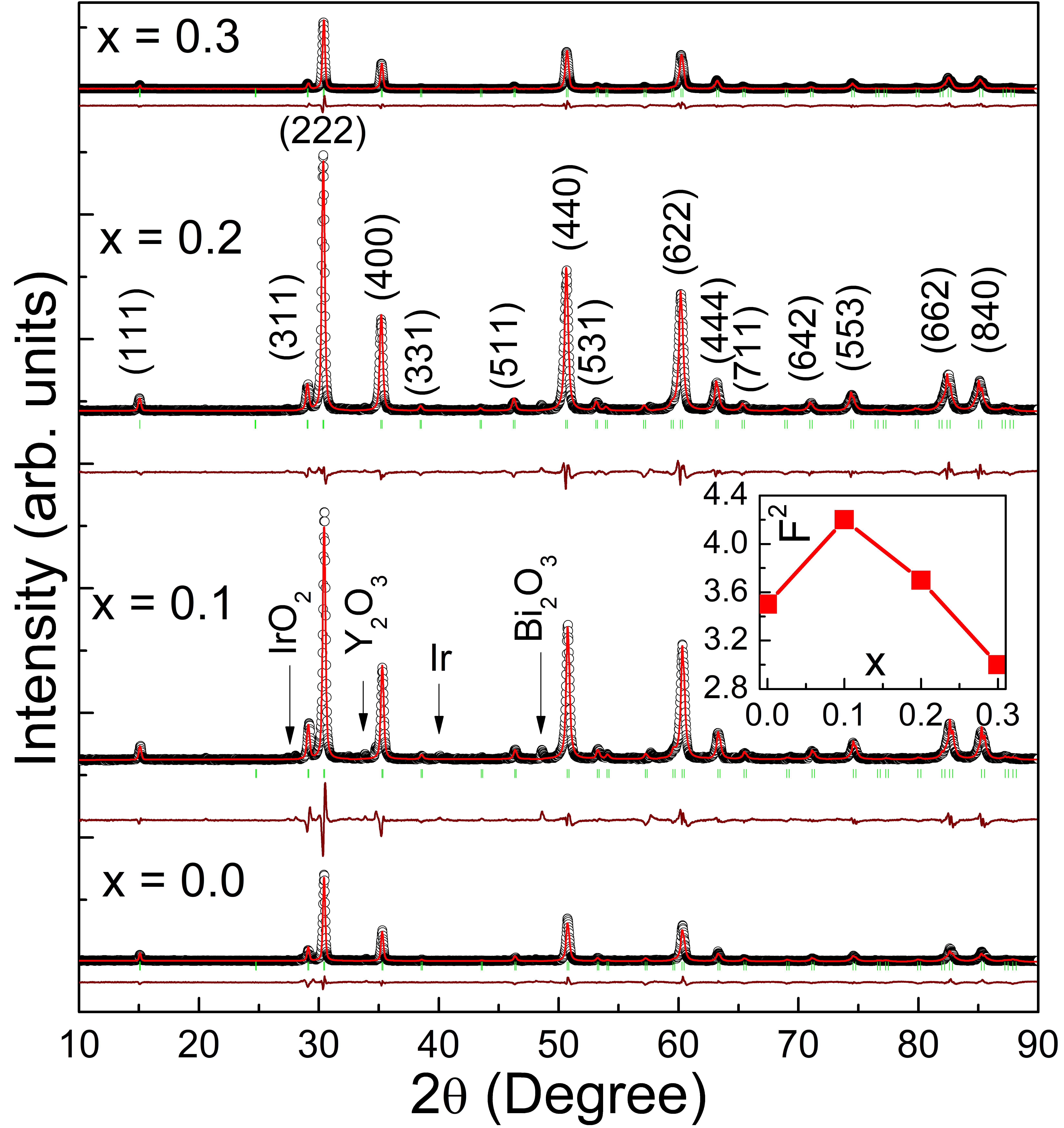}\\
	\caption{Powder XRD profile along with Rietveld refinement taking pyrochlore cubic structure have been shown for $Y_{2-x}Bi_xIr_2O_7$ series. Arrows represent impurity phases. Inset shows the $Bi$ doping concentration dependent goodness of fit value obtained from Rietveld fitting.}\label{fig:xrd}
\end{figure}

\begin{figure}
	\includegraphics[width=8.5cm]{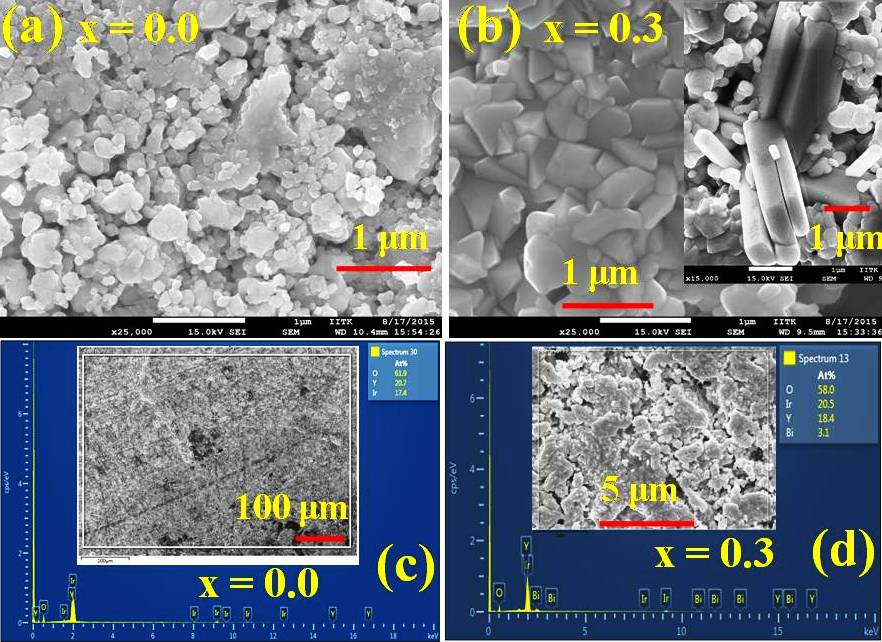}\\
	\caption{Field emission scanning electron microscope (FE-SEM) micrograph for polycrystalline (a) undoped (x = 0.0), and (b) x = 0.3 sample: inset shows formation of faceted single crystals. Energy dispersive x-ray spectrometry (EDX) spectra of (c) x = 0.0, and (d) x = 0.3 samples.}\label{fig:sem}
\end{figure}

\begin{figure}
	\includegraphics[width=8.5cm]{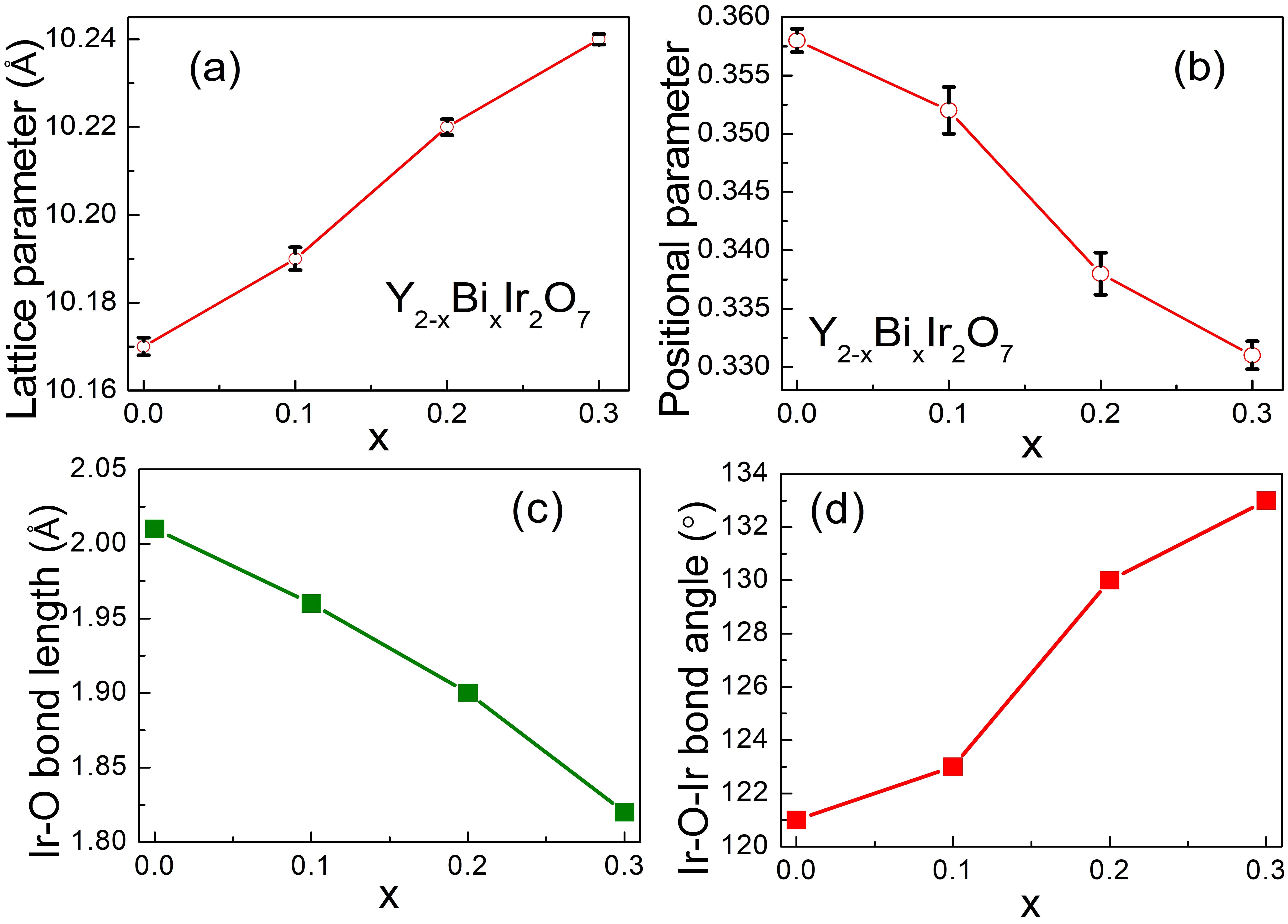}\\
	\caption{ $Bi$ doping content dependent (a) Lattice parameter, (b) positional parameters $r$ of $O$ atom, (c) $Ir-O$ bond length (d) $Ir-O-Ir$ bond angle for the $Y_{2-x}Bi_xIr_2O_7$ series.}\label{fig:bond}
\end{figure}

Figure~\ref{fig:xrd} shows powder X-ray diffraction (XRD) pattern for $Y_{2-x}Bi_xIr_2O_7$ series with x = 0.0, 0.1, 0.2 and 0.3. The XRD spectra with diffraction peaks for the undoped compound (x = 0.0) consistent with the reported study~\cite{Vinod1}. With substitution of $Bi$ at $Y$-site, major change in XRD peak position is not observed. It is expected because their matching ionic radii [$Bi^{3+}$ = 1.17$\AA$ and $Y^{3+}$ = 1.01$\AA$]~\cite{Shannon}. Indeed, XRD spectra does not change 	notably with doping concentration x except some impurities such as $Bi_2O_3$, $IrO_2$, $Y_2O_3$ and $Ir$, which is present minimally ($\leq 3\%$) in the series. The XRD spectra have been analyzed by Rietveld refinement program. The refinement results show all the compounds exihibt a nearly pure pyrochlore phase of cubic Fd$\bar{3}$m (227) except for few impurity phases shown in Fig.~\ref{fig:xrd} consistent with what others have been reported~\cite{Shapiro,Zhu}. These impuries are either diamagnetic or paramagnetic in nature, hence have negligible contribution to the physical properties. Statistical goodness of fit is defined as $F^2 = [R_{wp}/R_{exp}]^2$, where, $R_{exp}$ is the observed weighed profile factor and $R_{wp}$ the expected weighed profile factor in Rietveld refinement~\cite{Rietveld1}. The goodness of fit as a function of $Bi$ doping concentration (x) shown in inset of Fig.~\ref{fig:xrd} suggest the quality of fitting. Interestingly, Fig.~\ref{fig:bond}a shows that the lattice parameters increases with increasing $Bi$ doping concentration (x). It is expected that smaller $Y^{3+}$(1.01$\AA$) are replaced by larger $Bi^{3+}$(1.17$\AA$) ion in $Y_2Ir_2O_7$. The enhancement of lattice constant may be associated to the increase in $Y$-site ionic size. The representative field emission scanning electron microscope (FE-SEM) micrograph for two members of this present series i.e. x = 0.0 and 0.3 are shown in Fig.~\ref{fig:sem}a,b. The SEM micrograph of parent sample [Fig.~\ref{fig:sem}a] reveals nearly uniform distribution of closely packed grains of different shapes of particles with average particle size 1$\mu m$. On the other hand, Fig.~\ref{fig:sem}b shows faceted formation of particles. Inset of Fig.~\ref{fig:sem}b displays the faceted single crystal wires with typical length ranging between 1-50$\mu m$, loosely attached to the surface of polycrystalline sample of x = 0.3. Such formation of faceted single crystal possibly due to self-flux growth involving $Bi_2O_3$ as a flux~\cite{Aoyagi}. The chemical composition of desired element in polycrystalline $Y_{2-x}Bi_xIr_2O_7$ series was recorded by energy dispersive x-ray spectrometry (EDX) taken over a large area of the sample [shown in inset of Fig.~\ref{fig:sem}c,d], show the average expected presence of $Ir$, $Bi$, $Y$ and $O$ in nearly stoichiometric ratio as shown in Fig.~\ref{fig:sem}c,d. It can be noticed that the Ir/Y ratio is $\sim$ 0.84 in the x = 0.0 sample and the Ir/(Y+Bi) is $\sim$ 0.95 in the x = 0.3 sample, suggest that those samples are $Ir$-deficient. The low Ir/Y ratio for undoped compound is not surprising due to volatile nature of $IrO_3$~\cite{Matsuhira1,Yang1}. The higher value of Ir/(Y+Bi) ratio for x = 0.3 sample as compared to x = 0.0 might by associated to the volatile nature of Bi and stability of Ir in $Y_{2-x}Bi_xIr_2O_7$ series.

\begin{figure}
	\includegraphics[width=8.5cm]{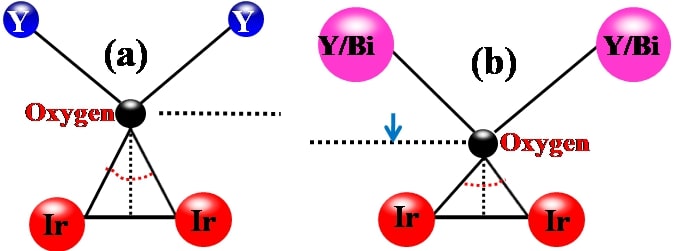}\\
	\caption{Oxygen coordination environment showing the effect of $R$ cation size on the local structure of (a) small $Y$ cation in $R_2Ir_2O_7$ system, (b) larger $Y$ cation in $Y_{2-x}Bi_xIr_2O_7$.}\label{fig:cryst}
\end{figure}

In ideal pyrochlore iridates with formula $R_2Ir_2O_6O'$, the four non-equivalent ions reside at following coordinates: $R$ at 16d site (0.5, 0.5, 0.5), $Ir$ at 16c (0, 0, 0) site, two types of oxygen, $O$ and $O'$ ions coordinated tetrahedrally reside at the 48f(r, 0.125, 0.125, where r termed as positional parameter) and 8b(0.375, 0.375, 0.375) sites, respectively. The Pyrochlore crystal structure show perfect octahedron about (0, 0, 0) for the value of x = 0.3125 give rise to perfect cubic field for $Ir$ cation in this $Ir-O_6$ octahedra~\cite{Gardner}. Figure~\ref{fig:bond}b reveals the adjustable positional parameter as a function of $Bi$ doping concentration x. The estimated value of the positional parameter for the parent compound turns out to be 0.355, greater than the ideal value, suggesting disordered and compressed $Ir-O_6$ octahedra. For the $Y_{2-x}Bi_xIr_2O_7$ series, positional parameter decreases with increase of the $Bi$ content, leading towards ideal $IrO_6$ octahedra with reduced crystal field. The doping dependence of $Ir-O$ bond distance and $Ir-O-Ir$ bond angle are shown in Fig.~\ref{fig:bond}c,d, respectively. The bond length decreases while bond angle increases with increasing concentration of $Bi$ in $Y_{2-x}Bi_xIr_2O_7$ series. This is due to the introduction of heavy $Bi^{3+}$ atom with greater ionic size than $Y^{3+}$ ~\cite{Koo,Bouchard}. This implies reduction of distortion in the $Ir-O_6$ octahedra and enhancement in the hybridization of $Y$($4p$)/$Bi$($6s$ and $6p$) with $O$($2p$) and $Ir$($5d$) states. This is also consistent with the non-monotonous variation of $F^2$ against Bi doping [inset of Fig.~\ref{fig:xrd}], giving rise to increased $A$-site ionic radius. The enhancement in $Ir-O-Ir$ bond angle and reduction in $Ir-O$ bond length can be understood by Fig.~\ref{fig:cryst}a,b. Cartoon shows the arrangement of atoms in the un-doped compound where the coordination environments of each oxygen atom is associated with two $Ir$ and two $Y$ cations to make a distorted octahedra along c-axis in the unit cell [Fig.~\ref{fig:cryst}a]. When $Bi^{3+}$ is introduced, it pushes the oxygen atom further away from the $R$-site along the perpendicular bisector of the $Ir-O-Ir$ triangle, thus increasing the $Ir-O-Ir$ bond angle [Fig.~\ref{fig:cryst}b].

\begin{figure}
	\includegraphics[width=8.5cm]{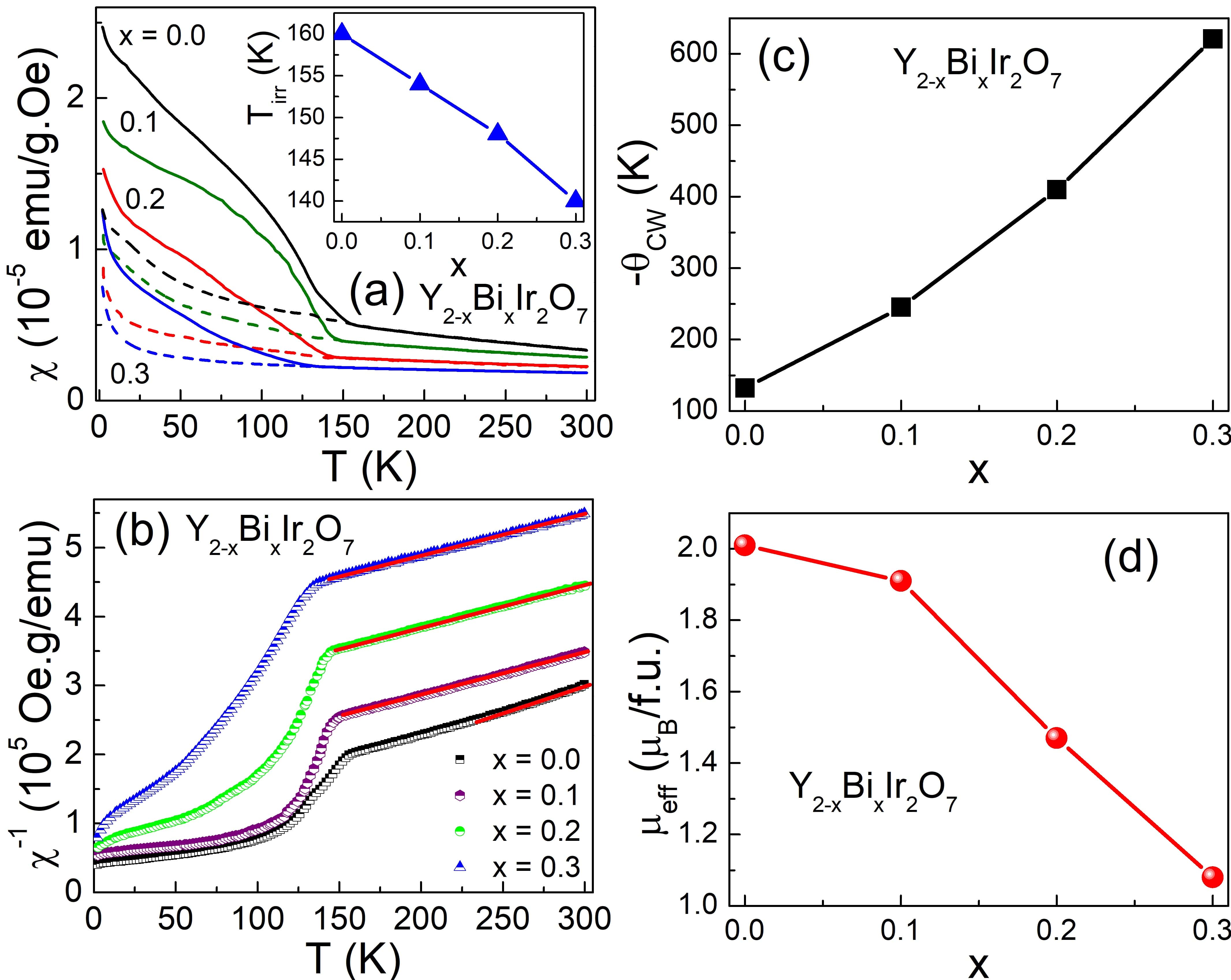}\\
	\caption{(a) Field cooled (solid continuous line) and zero field cooled (dashed line) magnetic susceptibilties as a function of temperature for $Y_{2-x}Bi_xIr_2O_7$ (x = 0.0, 0.1, 0.2. 0.3) series; inset shows $Bi$ doping concentration dependent magnetic irriversible temperature $T_{irr}$. (b) The inverse of magnetic susceptibility $\chi^{-1} = H/M$ as a function of temperature.  Red solid line associated to the the fit using the Curie-Weiss (CW) law. The variation of (c) Curie-Weiss temperature $\theta_{CW}$, and (d) effective magnetic moment $\mu_{eff}$ with $Bi$ doping content x.}\label{fig:mt}
\end{figure}

\begin{figure}
	\includegraphics[width=8.5cm]{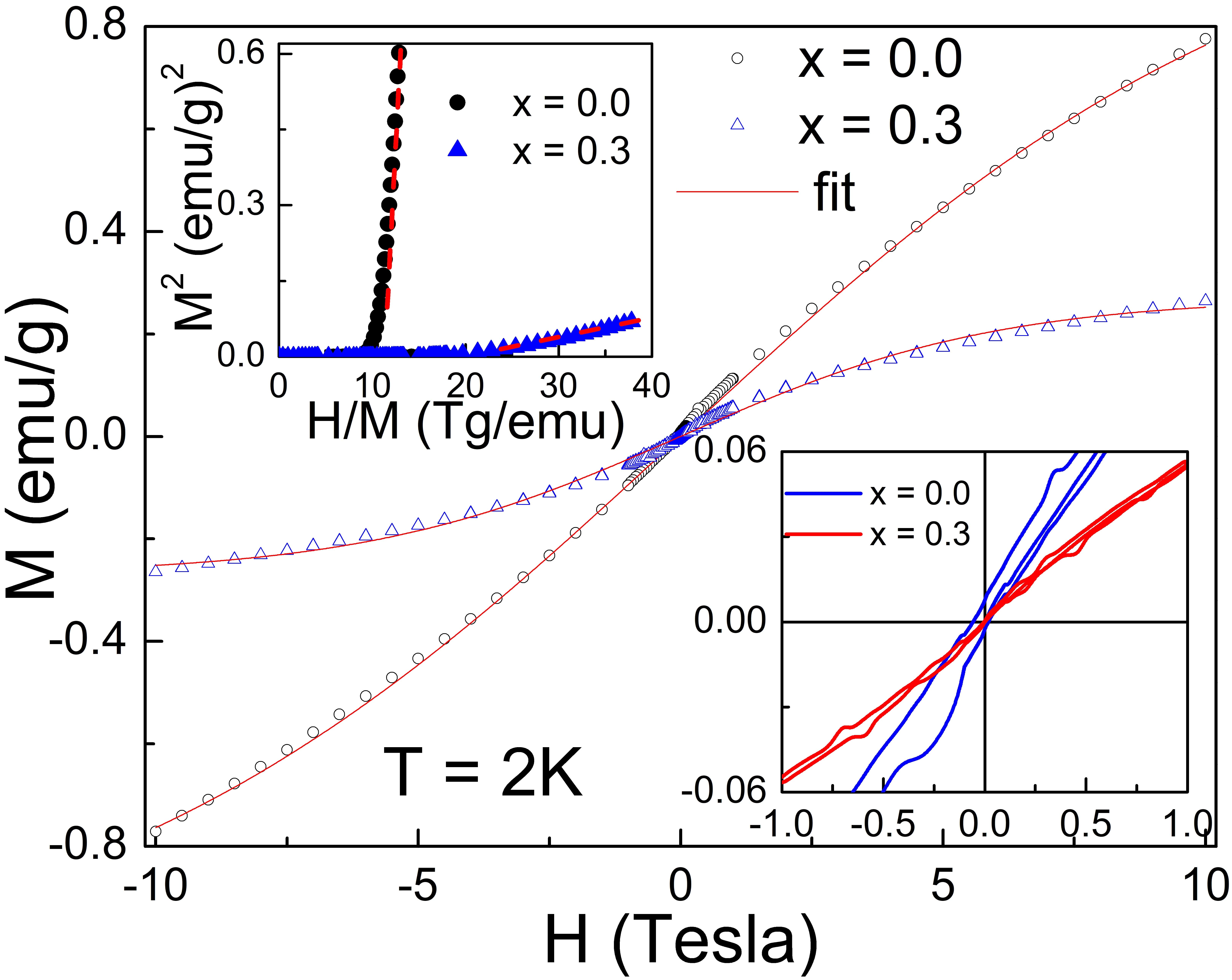}\\
	\caption{Isothermal magnetic hysteresis loops (M-H) curves measured at temperature 2K for two representative samples x = 0.0 and 0.3 of $Y_{2-x}Bi_xIr_2O_7$ series. The solid red line is the Brillouin function fit according to Eq.~\ref{eq:Brillouin}. Bottom inset shows the enlarged view of the low magnetic field and top inset displays conventional Arrott plot.}\label{fig:mh}
\end{figure}

The temperature dependence of the magnetic susceptibility $\chi = M/H$ for $Y_{2-x}Bi_xIr_2O_7$ (x = 0.0, 0.1, 0.2, 0.3) series measured at applied magnetic field 1kOe following the zero field cooled (ZFC) and field cooled (FC) protocol. For un-doped compound, it can be noticed that $\chi_{ZFC}$ and $\chi_{FC}$ branches contain a clear magnetic irreversibility around $T_{irr}$ $\sim$ 160K shown in Fig. 5a. Below $T_{irr}$ a large bifurcation between $\chi_{ZFC}$ and $\chi_{FC}$ branches is obtained. It can be noticed that $\chi_{ZFC}$ curve does not show any cusp at $T_{irr}$. The enhancement of susceptibility below the bifurcation temperature $T_{irr}$ suggest a bulk phase transition to an ordered state. Interestingly, the susceptibility enhances very sharply below T $\sim$ 25K, possibly due to emergence of very weak ferromagnetic components. This behaviour might be attributed to spin-glass like transition\cite{Vinod1,Harish1}. As shown in the Fig.~\ref{fig:mt}a, as Bi doping content increases: (I) the bifurcation temperature $T_{irr}$ shifts toward lower temperature, and (II) the difference between $\chi_{ZFC}$ and $\chi_{FC}$ branches decreases. Inset of Fig.~\ref{fig:mt}a shows almost linear reduction in $T_{irr}$ with Bi doping content. $\chi_{ZFC}$ and $\chi_{FC}$ branches entirely merges for fully Bi doped sample i.e. $Bi_2Ir_2O_7$~\cite{Bouchard,Baker,Qi}, imply paramagnetic-like behaviour. Figure~\ref{fig:mt}b shows the inverse of dc magnetic susceptibilty $\chi^{-1}$ vs T curve. Above the magnetic transition at high temperatures (140-300K), the linear part of the $\chi^{-1}(T)$ data for all the samples are analysed by fitting with the Curie-Weiss law, $\chi = \frac{N_A\mu_{eff}^2}{3k_B(T-\theta_{CW})}$, where $C$ and $\theta_{CW}$ are the Curie constant, Curie-Weiss temperature, $N_A$ is the Avogadro number and $k_B$ is the Boltzmann constant, respectively. We find negative value of $\theta_{CW}$ for $Y_{2-x}Bi_xIr_2O_7$ series suggests prodominantly antiferromagnetic (AFM) correlation. The absolute value of $\theta_{CW}$ is monotonically enhancing as more $Bi^{3+}$ ions are substituted into the $Y_2Ir_2O_7$ lattice shown in Fig.~\ref{fig:mt}c, suggest enhancement in antiferromagnetic (AFM) correlation. Due to the extended nature of electronic wave function of $Ir-5d$ orbital, the hybridization of $Ir-5d$ and $O-2p$ orbitals are strong enough supporting interatomic exchange interaction as discussed earlier [Fig.~\ref{fig:bond}b,c,d]. For parent compound (x = 0.0), the $Ir-O-Ir$ bond angle is 121$^0$ [very close to ideal value~\cite{Wan1,Wan2}], much larger than 90$^0$ promoting the super exchange interaction. With $Bi$ doping $Ir-O-Ir$ bond angle is enhancing and approaching towards 180$^0$. Therefore, $Ir-O-Ir$ AFM super exchange correlation is dominant in $Y_{2-x}Bi_xIr_2O_7$ series, lead to enhanced AFM correlation. It can be noticed that -$\theta_{CW}$ increases but the actual ordering temperature decreases. This suggests enhancement of frustration parameter [$f = \theta_{CW}/T_{irr}$] with $Bi$ doping. The effective magnetic moment $\mu_{eff}$ for the all samples plotted as a function of $Bi$ doping concentration shown in Fig.~\ref{fig:mt}(d). The estimated value of $\mu_{eff}$ for undoped compound is 2.2 $\mu_B$/f.u. which is larger than the expected Hund's rule value of 1.73 $\mu_B$/f.u. for $S = 1/2$. Similar inconsistency [i.e. obtained experimental value of $\mu_{eff}$ being larger than expected theoretical value of a spin 1/2] has also been reported~\cite{Harish1}. Such disagreement with Hund's rule value is not unusual in presence of crystal field effect and strong spin-orbit coupling.  It can be noticed that $\mu_{eff}$ decreases as non-magnetic $Bi$ content increases shown in Fig.~\ref{fig:mt}d. 

The magnetic field (H) dependence isothermal magnetization (M) for two representative samples x = 0.0, 0.3 of the $Y_{2-x}Bi_xIr_2O_7$ series measured at temperature 2K is shown in Fig.~\ref{fig:mh}. The M-H data show non-linear trends with no sign of saturation up to 10T. The coercive field $H_C$, saturation magnetization $M_S$ and remanent magnetization $M_R$ show considerable reduction compared to the un-doped sample. It is obvious that $Y_{2-x}Bi_xIr_2O_7$ series have tendency towards an increased AFM correlation, supports the enhanced negative $\theta_{CW}$ with increased Bi doping content. A narrower magnetic hysteresis loops for x = 0.3 compared to x = 0.0 sample can be seen at low magnetic field [bottom inset of Fig.~\ref{fig:mh}], suggesting very weak FM component. This result is possibly due to the canted antiferromagnetic spin coupling instead of ferromagnetic correlation. The coexistence of very weak ferromagnetic component on the large AFM background  suggests that only a partial fraction of short-range ordered spins freeze at low temperature into random direction. This behaviour might be attributed to spin-glass-like transition, possibly arising due to local structural disorder and magnetic frustration. Further, we studied the M-H data using Brillouin function fit~\cite{Blundell}. For $R_2Ir_2O_7$, the low temperature magnetic state can be explored including an effective spin J = 1/2, where

\begin{equation}\label{eq:Brillouin}
\frac{M}{M_S} = B_{1/2}(y)
\end{equation}

where $B_J(y) = \left[\frac{2J+1}{2J}\right]coth\left[\frac{y(2J+1)}{2J}\right] - \frac{1}{2J}coth\frac{y}{2J}$, $M_S = ng_J\mu_BJ$, y = $\left[\frac{g_J\mu_BJ}{k_BT}\right]H$, $\mu_B$ is Bohr magneton, $g_J$ is the Lande's g-factor. The value of effective magnetic moment $\mu_{eff}^{M(H)}$ is calculated using $\mu_{eff} = g_J\mu_B\sqrt{J(J+1)}$, where $g_J$ is obtained from the Brillouin function fit shown by solid red line in Fig.~\ref{fig:mh}. The estimated value of fitting parameters of M-H data are $g_J$ = 0.6 (x = 0.0), 1.1 (x = 0.3) and $\mu_{eff}^{M(H)}$ = 0.6$\mu_B$ (x = 0.0), 0.95$\mu_B$ (x = 0.3), respectively. While the estimated $\mu_{eff}^{M(H)}$ for x = 0.3 sample is very close to the effective magnetic moment $\mu_{eff}^{M(T)}$ = 1.1$\mu_B$ obtained from M-T data, the $\mu_{eff}^{M(H)}$ for x = 0.0 sample is smaller than $\mu_{eff}^{M(T)}$ = 2.2$\mu_B$. It is clear that the Brillouin function deviates from the experimental M-H data [Fig.~\ref{fig:mh}], suggesting non-paramagnetic trend. Top inset of Fig.~\ref{fig:mh} shows conventional Arrott plot~\cite{Harish2,Vinod6} for x = 0.0 and 0.3 samples. It is clear that estimated intercept [$\sim$ -2.1 (x = 0.0), -0.1 (x = 0.3)] from straight line fitting at $M^2$-axis is negative and decreasing with Bi doping content. It suggests that low temperature magnetic state is non-ferromagnetic type in un-doped compound, and Bi doping decreases the non-ferromagnetic type behaviour in $Y_{2-x}Bi_xIr_2O_7$ series.     

\begin{figure}
	\includegraphics[width=8.5cm]{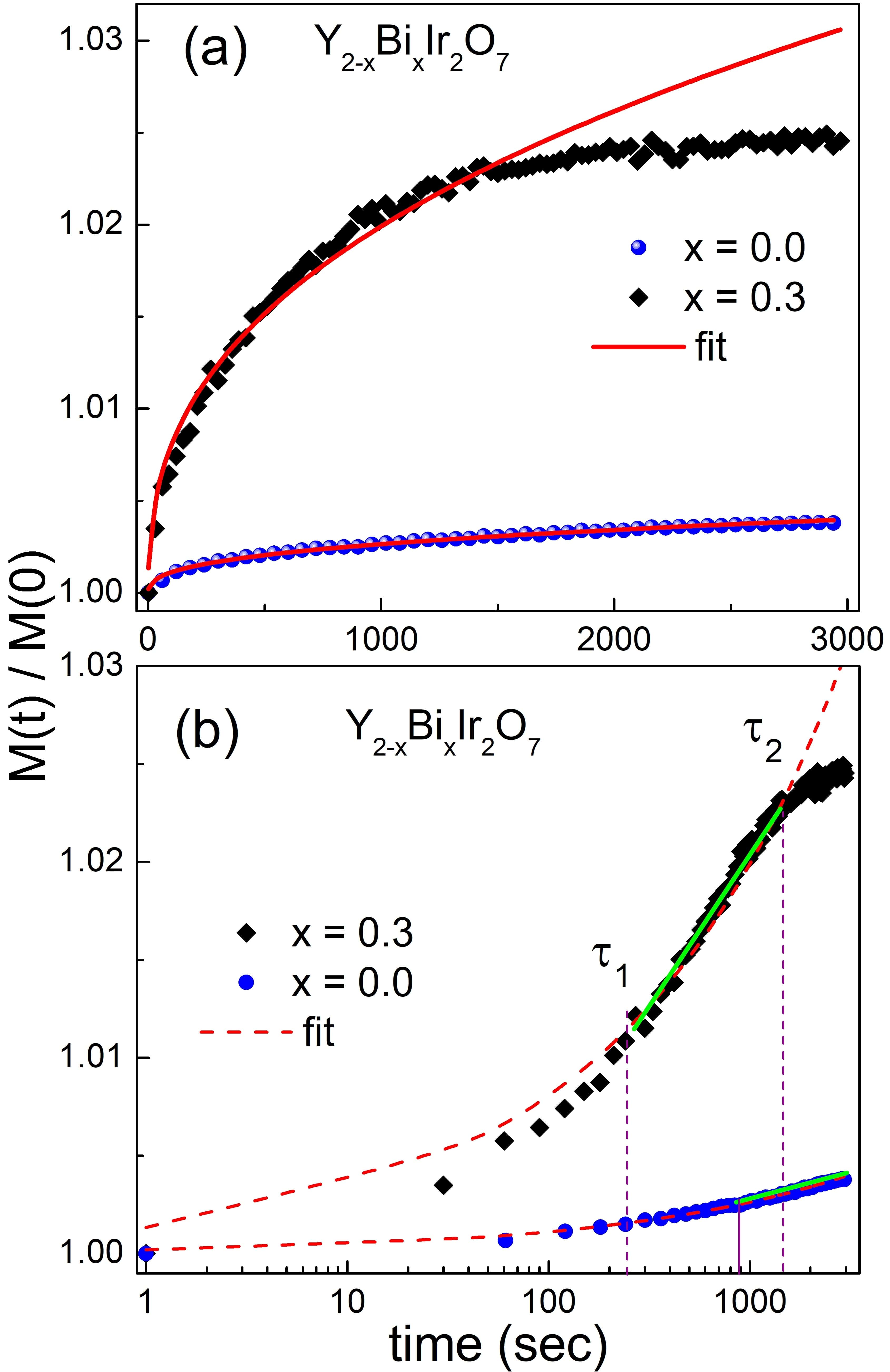}\\
	\caption{(a) Time dependence of the normalized isothermal remanent magnetization measured at temperature 5K with waited time $t_w$ = 10$^3$s for the two representative samples x = 0.0, 0.3 of the $Y_{2-x}Bi_xIr_2O_7$ series. Solid red line show corresponding fit of data using Eq.~\ref{eq:sg}. (b) Semi-log plot of normalized magnetization as a function of time. Dashed red line are due to the fit using Eq.~\ref{eq:sg}. Solid green line shows linear portion.}\label{fig:Relax}
\end{figure}

In order to further investigate the possibility of spin-glass-like behaviour in $Bi$ doped $Y_2Ir_2O_7$ pyrochlore iridates, we have measured the isothermal remanent magnetization. For this measurement, the samples are ZFC from room temperature to 5K. A dc magnetic field  of 1kOe is applied after stabilizing the temperature and waiting upto 10$^3$s, and magnetization is measured as a function of elapsed time t. Figure~\ref{fig:Relax}(a) shows the time dependence of ZFC isothermal remanent magnetization normalized with magnetization value M(t = 0) for two representative samples x = 0.0, 0.3 of the $Bi$ doped $Y_2Ir_2O_7$ series. For x = 0.0 sample, M(t)/M(0) increases with time without any sign of saturation, although x = 0.3 sample is trying to achieve saturation at higher time scale. We have analysed the normalized magnetic relaxation data with fitting of stretched exponential function~\cite{Sirena}
\begin{equation}\label{eq:sg}
\frac{M(t)}{M(t=0)} = exp\left[\frac{t}{\tau}\right]^\gamma
\end{equation}
where $\gamma$ is stretching exponent and $\tau$ is the characteristic relaxation time. The value of $\gamma$ give the information about the nature of energy barriers involved in magnetic relaxation. Syatem exhibiting a single energy barrier should give an exponential magnetic relaxation for the value of $\gamma \geq$ 1. In real system, several energy barriers involved in magnetic relaxation, lead to a distribution of relaxation times. The solid red lines shown in Fig.~\ref{fig:Relax} are due to fitting of stretched exponential Eq.~\ref{eq:sg}, giving the value of $\gamma$ = 0.37 (x = 0.0) and 0.4 (x = 0.3), respectively. Simultaneously, It can be noticed that $Bi$ doping decreases the relaxation time $\tau$ [x = 0.3, $\tau \sim 2.4\times 10^7$] almost by two order as compared to un-doped compound [x = 0.0, $\tau \sim 8.5\times 10^9$]. The estimated values of $\gamma$ and $\tau$ are in good agreement with the values observed for spin-glass system~\cite{Mydosh}. This clearly suggest that $Bi$ doping relaxes the system at a faster rate. Figure~\ref{fig:Relax}(b) displays time dependent semi-logarithmic plot of normalized magnetic relaxation data. Parent compound shows continuous increase of magnetization with time. On the other hand, sample x = 0.3 enters in a regime of saturation at higher time scale (after $\tau_2$). In this regime, the magnetization change tends to zero for $t \gg \tau_2$ and for $t \ll \tau_1$, the magnetization varies linearly with time~\cite{Sirena,Yoo}, consistent with a distribution function of finite width. Therefore, it can be concluded that $Y_{2-x}Bi_xIr_2O_7$ series looks like glassy mixed phase.

\begin{figure}
	\includegraphics[width=8.5cm]{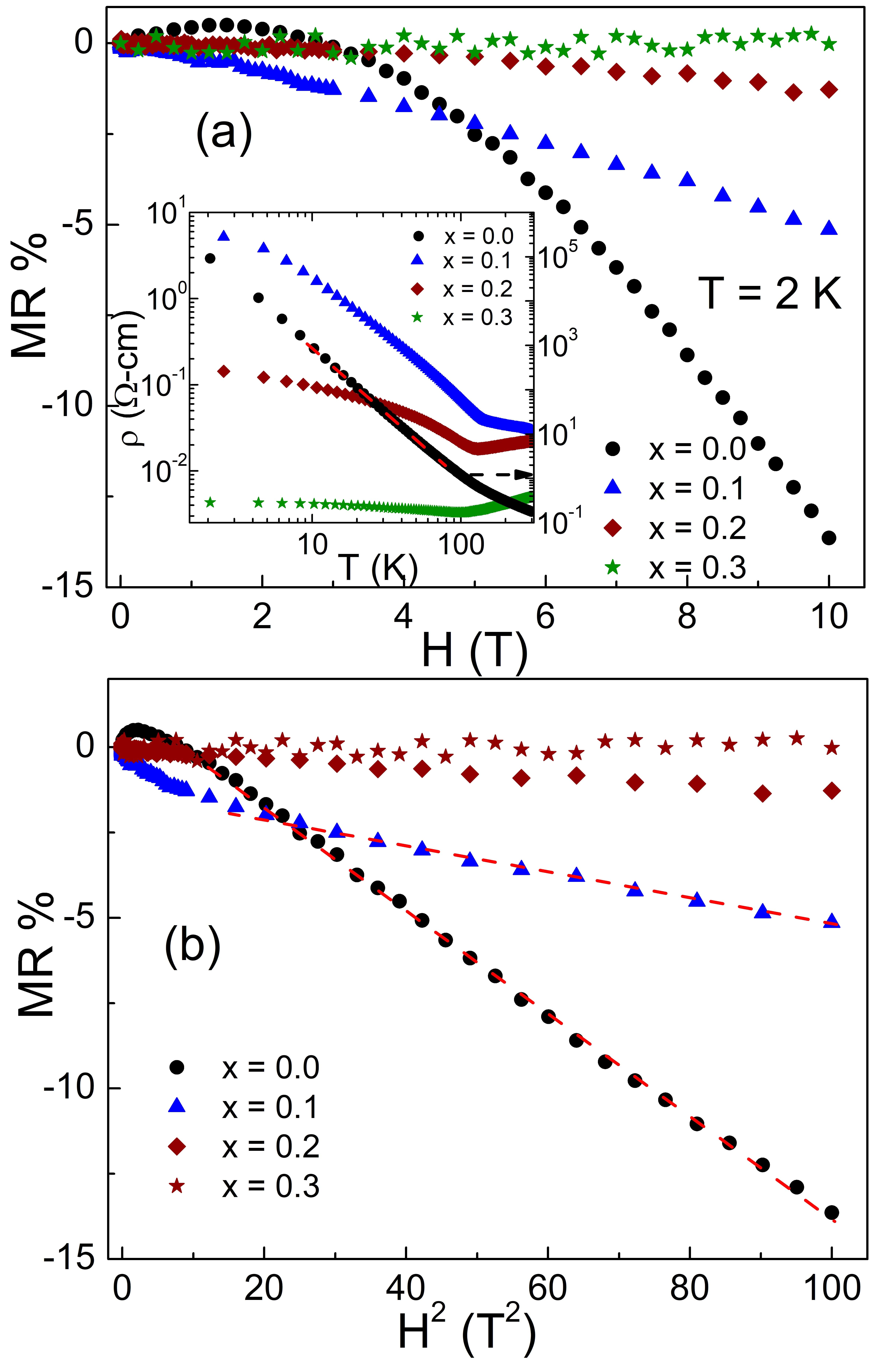}\\
	\caption{(a) Magnetoresistance as function of applied magnetic field for $Y_{2-x}Bi_xIr_2O_7$ series measured at temperature 2K. Inset shows log-log plot of resistivity ($\rho$) as a function of temperature (T); $\rho$(T) of x = 0.0 sample is shown on right y-axis. The solid black line represnts the theoretical fit due to power law. (b) Quadratic field dependence of MR.}\label{fig:RT}
\end{figure}

Electrical resistivity as a function of temperature for $Y_{2-x}Bi_xIr_2O_7$ series is shown in inset of Fig.~\ref{fig:RT}a. The value of resistivities at room temperature for all the samples are as follows: 0.18 $\Omega-cm$ (x = 0.0), 0.03 $\Omega-cm$ (x = 0.1), 0.02 $\Omega-cm$ (x = 0.2), and 0.005 $\Omega-cm$ (x = 0.3). The x = 0.0 sample is known to show insulating like behaviour in the whole temperature range~\cite{Vinod1,Vinod3}, which diminishes with $Bi$ doping. For doping level x = 0.1, insulator like temperature dependence is observed, while for x = 0.2 and 0.3, we find much lower resistivity with minima at temperatures $\sim$ 140 K, 130 K, respectively. As the conduction mechanism in parent sample follows the power law model, $\rho = \rho_0T^{-n}$, where n is the power law exponent \& $\rho_0$ is the prefactor, respectively, the $\rho(T)$ data are analyzed by fitting to the power law in the temperature range 14-90K (x = 0.0), 14-70K (x = 0.1), 14-50K (x = 0.2), 14-35K (x = 0.3) shown in inset of Fig.~\ref{fig:RT}a. The fitting parameters are estimated to be n, \& $\rho_0$  $\sim$ 3.1 \& 9$\times$10$^5$ $\Omega-cm$ (x = 0.0), 1.4 \& 42 $\Omega-cm$ (x = 0.1), 0.6 \& 0.3 $\Omega-cm$ (x = 0.2), 0.1 \& 0.005 $\Omega-cm$ (x = 0.3), samples. The MIT state in $Y_{2-x}Bi_xIr_2O_7$ series is of general interest. While replacement of $Y^{3+}$ with $Ca^{2+}$ leads to hole doping, which in turn, gives rise to a finite density of states near the Fermi level~\cite{Fukazawa,Zhu}, the introduction of $Bi$ in $Y_2Ir_2O_7$ system does not create holes but fills up the $5d$ orbital. The greater ionic radius of $Bi^{3+}$ (1.17 $\AA$) compared to $Y^{3+}$ (1.01 $\AA$) introduces disorder into the compound by spatially deforming the cell volume. Substitution of $Bi^{3+}$ ions enhances the hybridization between the $Bi$ $6s/6p$ orbital with the $Ir$ $5d$ orbital~\cite{Koo,Bouchard}. The energy of this hybridization is much larger than the SOC and coulomb correlation, which makes the $Ir$ $5d$ bandwidth wider and drives the system into a metallic state. However, another fact that may push $Y_{2-x}Bi_xIr_2O_7$ to MIT state is an enhancement in the $Ir-O-Ir$ bond angle as $Bi$ doping content increases. It is known that for strong hybridization of $p$ orbital of oxygen with $t_{2g}$-$Ir^{4+}(5d^5)$ orbital, $Ir-O-Ir$ angle should be 180$^0$ in a $\Pi$ type interaction. Because of the enlarged nature of $Ir-5d$ orbitals, the mixing of $O(2p)-Ir(5d)$ is sufficient, thus favouring the interatomic exchange interaction.  The reported $Ir-O-Ir$ bond angle is 116$^0$ in $Y_2Ir_2O_7$ system, much greater than 90$^0$ supports the $Ir-O-Ir$ antiferromagnetic superexchange interaction, leading towards a noncollinear and strong magnetic frustration~\cite{Harish1}. On the other hand, the $Ir-O-Ir$ angle is 131.4$^0$ in metallic $Bi_2Ir_2O_7$~\cite{Kennedy}. Thus, it could be argued that the reason for improved conduction in $Bi$ doped $Y_2Ir_2O_7$ compounds, possibly due to enhancement in $Ir-O-Ir$ bond angle as shown in Fig.~\ref{fig:bond}d. Strikingly, such dependency of electrical transport behaviour on electronic structure parameters is almost similar to the behaviour under the applied hydrostatic pressure for the compound $A_2Ir_2O_7$ (A = Gd, Eu, Sm and Bi)~\cite{Wei,Wei1}. 

Figure~\ref{fig:RT}a shows the magnetoresistance (MR) as a function of magnetic field measured at temperature 2K. MR is defined as $\left[MR = \frac{\rho(H)-\rho(0)}{\rho(0)}\times 100\right]$. The undoped compound (x = 0.0) is known to show  positive MR at low magnetic field accompanied by negative MR with quadratic field dependence at higher magnetic field~\cite{Vinod3}. The negative MR is progressively suppressed with Bi doping shown in Fig.~\ref{fig:RT}a, although, quadratic field dependence is still observed shown in Fig.~\ref{fig:RT}b. The MR behaviour produced by quantum interference in presence of strong SOC is described by two time scales~\cite{Kallaher}. (I) spin-flip time $\tau_{so}$, and (II) the dephasing time $\tau_\phi$. When $\tau_\phi \gg \tau_{so}$, i.e. non-magnetic impurity scattering is dominant, destructive quantum interference between time reversed trajectories is obtained, leading to weak antilocalization and a positive MR. In $\tau_\phi < \tau_{so}$ regime, i.e. when SOC is weak, the quantum interference is always in-phase and constructive giving rise to negative MR with increasing magnetic field (Weak localization). The weakening of negative MR with $Bi$ doping imply enhancement of spin-orbit coupling due to the replacement of lighter $Y^{3+}$ with non-magnetic heavier $Bi^{3+}$.

\begin{figure}
	\includegraphics[width=8.5cm]{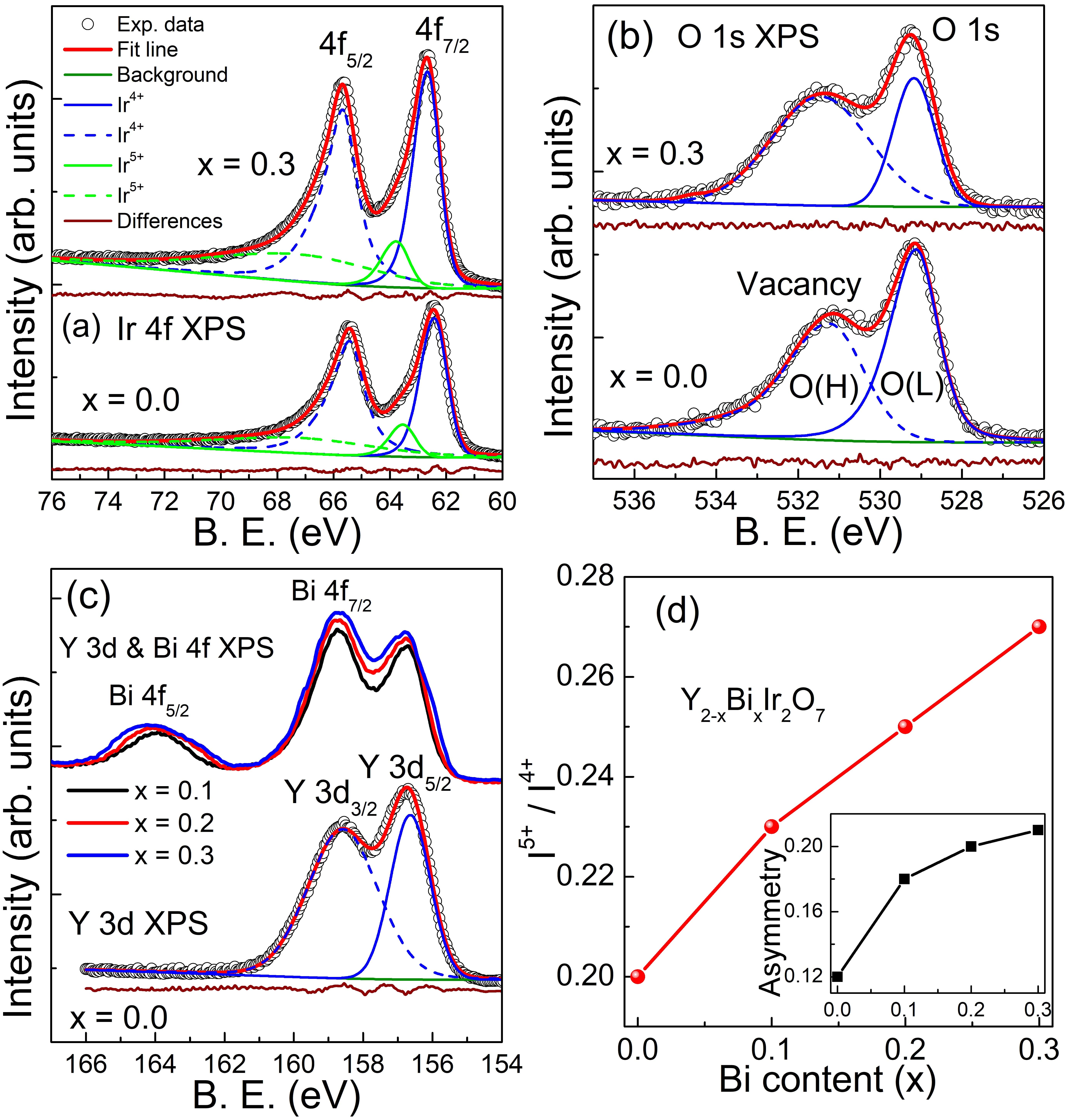}\\
	\caption{(a) deconvoluted $Ir$ 4f peaks for x = 0.0, 0.3 samples, (b) deconvoluted $O$ 1s peaks of the x = 0.0, 0.3 compounds. (c) $Y$ 3d and $Bi$ 4f XPS for x = 0.0, 0.1, 0.2, 0.3 compounds, (d) Intensity ratio $I^{5+}$/$I^{4+}$ $Ir^{5+}$ and asymmetry factor (shown in inset) as a function of $Bi$ doping content (x), where I = Ir.}\label{fig:XPS}
\end{figure}

\begin{figure}
	\includegraphics[width=8.5cm]{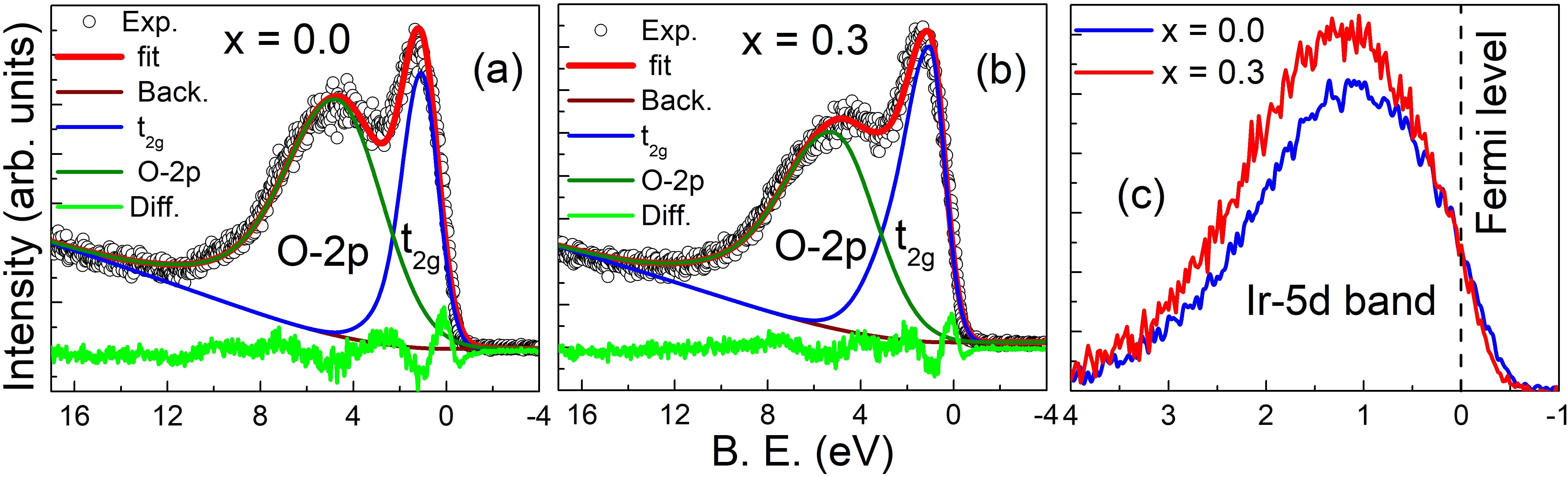}\\
	\caption{ Deconvoluted XPS valence band spectrum of Ir-5d (a) x = 0.0, (b) x = 0.3 samples. (c) Density of states estimated according to the protocol reported elsewhere~\cite{Singh}.}\label{fig:VBS}
\end{figure}

Figure~\ref{fig:XPS}a shows de-convoluted spectra of $Ir$ 4f core-level XPS of two representative sample x = 0.0, 0.3 of $Y_{2-x}Bi_xIr_2O_7$ series using asymmetric Gauss-Lorentz sum function. XPS peaks are labeled following the protocol reported elsewhere~\cite{Vinod1,Vinod2,Yang1,Yang2}. For $Ir^{4+}$, 4$f_{7/2}$ and 4$f_{5/2}$ electronic states arise at binding energy around 62eV and 65eV due to spin-orbit coupling, respectively, which are represented by blue solid line and blue dashed line in Fig~\ref{fig:XPS}a. Similarly, higher oxidation state $Ir^{5+}$ centered at 63.7eV and 67eV shown by green solid line and green dashed line, respectively in Fig~\ref{fig:XPS}a for 4$f_{7/2}$ and 4$f_{5/2}$. Fitted result show very small contribution from the $Ir^{5+}$ oxidation state along with major contribution from the $Ir^{4+}$ charge state for x = 0.0 sample, consistent with previous reports~\cite{Zhu,Vinod1,Vinod3,Vinod5}. For $Bi$ doped compounds, the contribution from $Ir^{5+}$ is minimally increased. It suggest the marginal increase in $Ir$ oxidation state, i.e. $Ir^{4+}$ and $Ir^{4.2+}$, as $Bi$ doping content increases shown in Fig.~\ref{fig:XPS}d. Therefore, it seems that replacement of $Y^{3+}$ by $Bi^{3+}$ does not support the formation of higher oxidation states. However, it can create anion vacancies at $O'$ sites which is favorable due to presence of polarizable $Bi^{3+}$ cation at $Y^{3+}$ site~\cite{Kennedy}. Figure~\ref{fig:XPS}c shows $Y3d$ XPS for x = 0.0, 0.1, 0.2, 0.3 samples. Variation in the $Bi-4f$ and $Y-3d$ XPS spectra with different doping concentration of $Bi$ can also be seen. The $Y-3d$ shows a single feature at 156.4eV and 158.3eV, suggest that only $Y^{3+}$ is present in the x = 0.0 sample. On the other hand, we observe a superposition of the $Y-3d$ and $Bi-4f$ XPS peaks. It is clear that $Y-3d$ peaks exhibit the $Bi-4f$ peak due to overlapping of energy ranges. The details about XPS of $Bi$ doped $Y_2O_3$ is given eleswhere~\cite{Jafer}. In further support of mixed oxidation state, the $O$ 1s XPS spectra for x = 0.0, 0.3 compounds are shown in Fig.~\ref{fig:XPS}b. The peaks located at binding energies 529eV and 531.4eV are attributed as lower $O(L)$ and higher $O(H)$ binding energy peaks, respectively. The $O(L)$ is assigned to oxygen lattice of $Y_{2-x}Bi_xIr_2O_7$ series while $O(H)$ is related to oxygen vacancies or defects~\cite{Xinyu}. The enhancement in the peak area ratio of $\frac{O(H)}{O(L)}$ against $Bi$ doping content suggests the enhancement in oxygen vacancies in the compounds with increased $Bi$ doping. We have estimated asymmetry factor in $Ir$ 4f XPS following the protocol reported eleswhere~\cite{Kennedy,Yang2} and plotted as a function of $Bi$ doping content (x) shown in inset of Fig.~\ref{fig:XPS}. It is obvious that less conducting parent sample exhibit less asymmetry compared to high conducting x = 0.1, 0.2, 0.3 samples. Asymmetry in $Ir$ 4f line shapes can be attributed to the $5d$ conduction electron screening, which could also explain the enhanced conductivity of the doped samples.

Furthere, we measured XPS valence-band spectra (VBS) to determine the unoccupied and occupied density of states (DOS) for x = 0.0 and 0.3 samples. We have deconvoluted the XPS valence-band spectra using a sum of Lorentzians and Gaussian to determine the contributions of O-2p~\cite{Singh}. The two peaks, centered at binding energy 1.2eV and 4.7eV shown in Fig.~\ref{fig:VBS}a,b are assigned as Ir-5d ($t_{2g}$) state and O-2p state, respectively. The contribution of Ir-5d band is estimated by subtracting the O-2p contributions shown in Fig.~\ref{fig:VBS}c. We obtain enhancement in DOS attributed with Ir-5d band in x = 0.3 sample as compared to un-doped compound, which is consistent with lower resistivity and MIT behaviour of the x = 0.3 sample. The enhancement of Ir-5d peak close to Fermi energy level ($E_F$) regime in x = 0.3 sample might be due to the marginally enhanced oxidation state of Ir compared to x = 0.0 compound. This makes the $Y_{2-x}Bi_xIr_2O_7$ system more conducting. 

\begin{figure}
	\includegraphics[width=8.5cm]{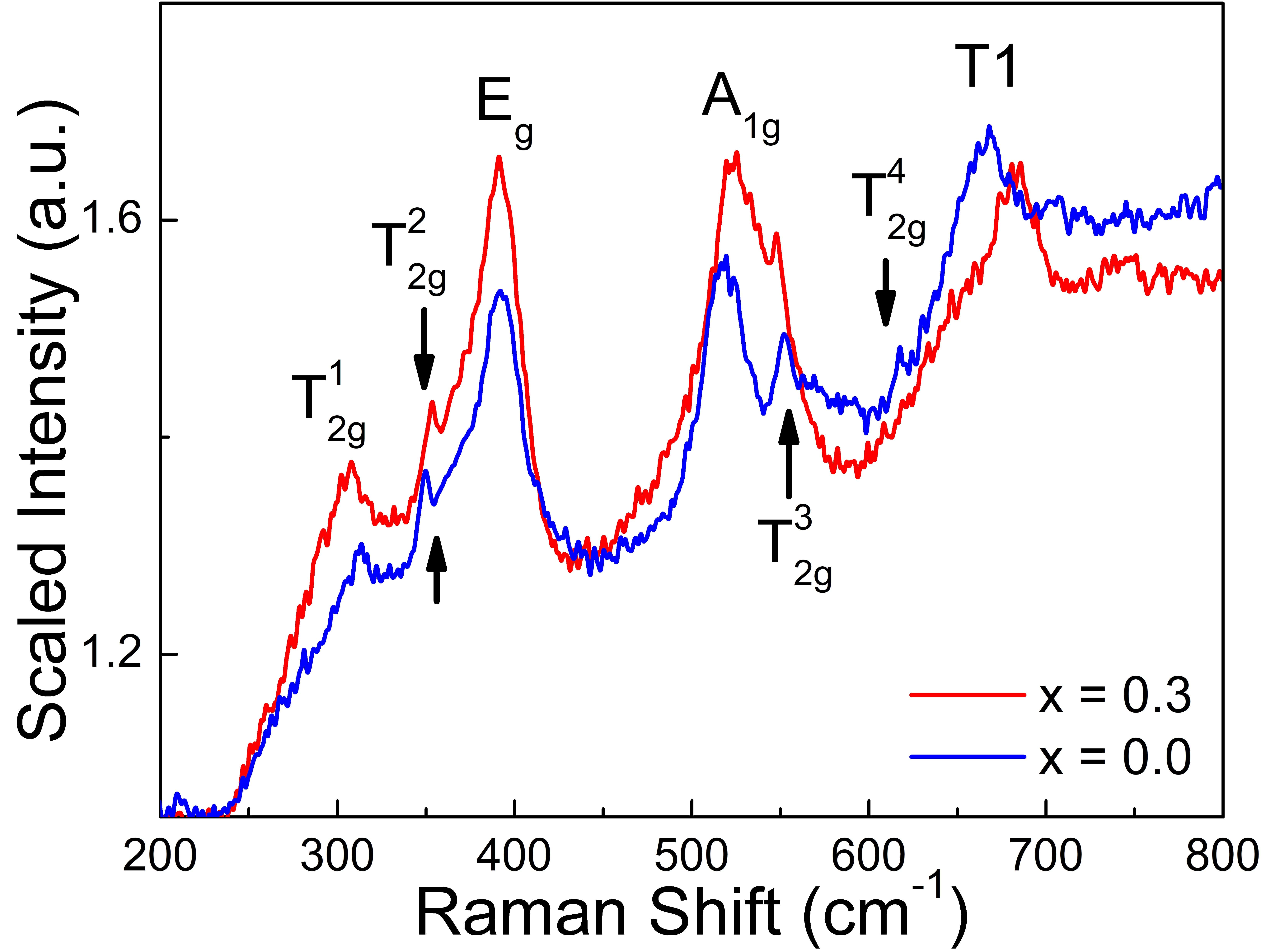}\\
	\caption{Intensity normalized Raman spectra of two representative samples x = 0.0, 0.3 of $Y_{2-x}Bi_xIr_2O_7$ series measured at room temperature in the spectral range 200-800cm$^{-1}$.}\label{fig:Raman}
\end{figure}

\begin{table}\label{table:Raman}
	\caption{Indexed Raman peaks with their respective wavenumbers}
	\begin{tabular}{|c|c|c|c|c|c|c|c|c|}
		\hline
		Modes & $T^1_{2g}$ & $T^2_{2g}$ & $T^3_{2g}$ & $T^4_{2g}$ & $E_g$ & $A_{1g}$ & $T1$\\
		\hline 
		Wavenumber & 305 & 350 & 550 & 630 & 391 & 523 & 665\\
		\hline 
	\end{tabular}
\end{table}

To further explore structural change and the possible cause behind the huge reduction in electrical resistivity in $Bi$ doped samples as compared to parent compound, Raman spectroscopy has been carried out. Figure~\ref{fig:Raman} shows Raman spectra recorded at room temperature for x = 0.0, 0.3 samples. An ideal cubic pyrochlore iridate with formula $R_2Ir_2O_6O'$, space group $Fd\bar{3}m$ should have six Raman active fundamental modes~\cite{Harish3} distributed among $A_{1g}$ + $E_g$ + $4T_{2g}$ modes, involving only vibrations of oxygens at both $48f$ and $8b$ sites. Table I show indexed modes with their respective wave numbers. The four peaks named as $T^1_{2g}$, $T^2_{2g}$, $T^3_{2g}$ and $T^4_{2g}$ are Raman active while the peak named as $T1$ has been reported as second order scattering for pyrochlore oxides. The most characteristic features of Ir-pyrochlore spectra are a sharp mode at $\sim$ 391cm$^{-1}$, and $\sim$ 523cm$^{-1}$ which includes contributions from the $E_g$ mode and $A_{1g}$ mode, respectively. It have been shown that intermediate vibrational modes $E_g$-$T^3_{2g}$ and $A_{1g}$-$T^2_{2g}$ arise due to vibrations of $Ir-O$ and $R-O'$ bond. Here, $A_{1g}$ mode is related to the positional parameter, directly affecting the $R-O'$ bond environment. For x = 0.3 sample, $E_g$-$T^3_{2g}$ vibrational mode shift towards lower wave number, while $A_{1g}$-$T^2_{2g}$ modes towards higher wave number as compared to parent compound. This feature are arising due to enhancement in hybridization between $Y(4p)/Bi(6s)$, $O-2p$ and $Ir-5d$ orbitals suggest reduction in $Ir-O$ bonds consistent with XRD shown in Fig.~\ref{fig:bond}c and enhancement in $R-O'$ bonds. In this process $t_{2g}$ bandwidth increases as the $R$-site ionic radius increases lead to enhanced electrical conductivity. 

The plausible scenario for the MIT in the $Y_{2-x}Bi_xIr_2O_7$ series compounds is related to the modification in structural parameters, arising due to the isovalent $Bi$ substitution at the $Y$-site as distinct from the previous studies which rarely focus on the relation between structural parameters and the magnetic and the electrical transport properties~\cite{Aito,Soda,Zhu,Fukazawa}.

\section{Conclusion}
	We have investigated the structural, magnetic, and electrical transport properties of the pyrochlore iridates $Y_{2-x}Bi_xIr_2O_7$. The replacement of $Y^{3+}$ with $Bi^{3+}$ stabilizes the antiferromagnetic correlation and enhances electronic conductivity. The XRD analysis shows $Ir-O-Ir$ bond angle increases and $Ir-O$ bond length decreases as distortion in $Ir-O_6$ octahedra is reduced against increasing $Bi^{3+}$ doping content. The X-ray photoemission spectroscopy measurements suggest marginal enhancement in $Ir$ oxidation states ranging from $Ir^{4+}$ to $Ir^{4.2+}$ with $Bi$ doping, promoting the anion vacancies at $O'$ site. The enhancement in electrical conductivity with $Bi^{3+}$ is likely a consequence of hybridization between the $Y(4p)/Bi(6s)$, $O-2p$ and $Ir-5d$ orbitals. Raman spectroscopy shows contraction in $Ir-O$ bonds [consistent with XRD results] and elongation in $R-O'$ bonds as disorder and phononic oscillation are reduces by $Bi$ doping, leading to orders of magnitude enhancement of electrical conductivity along with MIT. On the other hand, enhancement in antiferromagnetic correlation can be explained in term of increased $Ir-O-Ir$ bond angle against $Bi$ doping. Bifurcation between $\chi_{ZFC}$ and $\chi_{FC}$ at higher temperature with hysteretic isothermal magnetization at low magnetic field might be associated to spin-glass-like transition rather than long-range ordering, as confirmed by magnetic relaxation measurements.


\begin{thebibliography}{99}

\bibitem{Krempa1} William Witczak-Krempa, Gang Chen, Yong Baek Kim, and Leon Balents, \textit{Annu. Rev. Condens. Matter Phys.} {\bf{5}}, 57-82 (2014).

\bibitem{Pesin} Dmytro Pesin and Leon Balents, \textit{Nat. Phys.} {\bf{6}}, 376-381 (2010).

\bibitem{Wan1} X. Wan, Ari M. Turner, Ashvin Vishwanath, and Sergey Y. Savrasov, \textit{Phys. Rev. B} {\bf{83}}, 205101 (2011).

\bibitem{Hanawa} M. Hanawa, Y. Muraoka, T. Tayama, T. Sakakibara, J. Yamaura, and Z. Hiroi, \textit{Phys. Rev. Lett.} {\bf{87}}, 187001 (2001).

\bibitem{Mandrus}  D. Mandrus, J. R. Thompson, R. Gaal, L. Forro, J. C. Bryan, B. C. Chakoumakos, L. M. Woods, B. C. Sales, R. S. Fishman, and V. Keppens, \textit{Phys. Rev. B} {\bf{63}}, 195104 (2001).

\bibitem{Koo} H.-J. Koo, M.-H. Whangbo and B. J. Kennedy, \textit{J. Solid State Chem.} {\bf{136}}, 269-273 (1998).

\bibitem{Harish1} Harish Kumar, R. S. Dhaka, and A. K. Pramanik, \textit{Phys. Rev. B} {\bf{95}}, 054415 (2017).

\bibitem{Harish2} Harish Kumar and A. K. Pramanik, \textit{J. Phys. Chem. C} {\bf{123}}, 13036-13046 (2019).

\bibitem{Vinod1} Vinod Kumar Dwivedi, and Soumik Mukhopadhyay, \textit{J. Appl. Phys.} {\bf{125}}, 223901 (2019).

\bibitem{Vinod2} Vinod Kumar Dwivedi, and Soumik Mukhopadhyay, \textit{AIP Conf. Proc.} {\bf{1953}}, 120035 (2018).

\bibitem{Hui1} Hui Liu, Jian Bian, Shiyun Chen, Yu Wang, Yuan Feng, Wei Tong, Yu Xie, Baolong Fang, \textit{Physica B: Condensed Matter} {\bf{568}}, 60-65 (2019).

\bibitem{Tafti} F. F. Tafti, J. J. Ishikawa, A. McCollam, S. Nakatsuji, and S. R. Julian, \textit{Phys. Rev. B} {\bf{85}}, 205104 (2012).

\bibitem{Wei} Wei Liu, Hui Han, Langsheng Ling, Long Ma, Li Pi, Lei Zhang, Yuheng Zhang, \textit{Ceram. Int.} {\bf{43}}, 17100 (2017).

\bibitem{Wei1} Wei Liu, Hui Han, Long Ma, Li Pi, Lei Zhang, Yuheng Zhang, \textit{J. Alloys Compd.} {\bf{741}}, 182 (2018).

\bibitem{Vinod3} Vinod Kumar Dwivedi, Abhishek Juyal and Soumik Mukhopadhyay, \textit{Mater. Res. Express} {\bf{3}}, 115020 (2016).

\bibitem{Vinod4} Vinod Kumar Dwivedi, and Soumik Mukhopadhyay, \textit{AIP Conf. Proc.} {\bf{1665}}, 050160 (2015).

\bibitem{Vinod5} Vinod Kumar Dwivedi, and Soumik Mukhopadhyay, \textit{J. Magn. Magn. Mater} {\bf{484}}, 313-319 (2019).

\bibitem{Vinod6} Vinod Kumar Dwivedi, and Soumik Mukhopadhyay, \textit{Physica B: Condensed Matter} {\bf{571}}, 137-141 (2019).

\bibitem{Matsuhira1} K. Matsuhira, M. Wakeshima, Y. Hinatsu, and S. Takagi, \textit{J. Phys. Soc. Jpn.} {\bf{80}}, 094701 (2011).

\bibitem{Yanagishima} Daiki Yanagishima, and Yoshiteru Maeno, \textit{J. Phys. Soc. Jpn.} {\bf{70}}, 2880-2883 (2001).

\bibitem{Taira} N. Taira, M. Wakeshima, and Y. Hinatsu, \textit{J. Phys.: Condens. Matter} {\bf{13}}, 5527 (2001).

\bibitem{Zhu} W. K. Zhu, M. Wang, B. Seradjeh, Fengyuan Yang, and S. X. Zhang, \textit{Phys. Rev. B} {\bf{90}}, 054419 (2014).

\bibitem{Fukazawa} H. Fukazawa and Y. Maeno, \textit{J. Phys. Soc. Jpn.} {\bf{71}}, 2578 (2002).

\bibitem{Shapiro} M. C. Shapiro, Scott C. Riggs, M. B. Stone, C. R. de la Cruz, S. Chi, A. A. Podlesnyak, and I. R. Fisher, \textit{Phys. Rev. B} {\bf{85}}, 214434 (2012).

\bibitem{Hui2} Hui Liu, Wei Tong, Langsheng Ling, Shile Zhang, Ranran Zhang, Lei Zhang, Li Pi, Changjin Zhang, and Yuheng Zhang, \textit{Solid State Commun.} {\bf{179}}, 1-5 (2014).

\bibitem{Aito} Natsuki Aito, Minoru Soda, Yoshiaki Kobayashi, and Masatoshi Sato, \textit{J. Phys. Soc. Jpn.} {\bf{72}}, 1226 (2003).

\bibitem{Soda} M. Soda, N. Aito, Y. Kurahashi, Y. Kobayashi, M. Sato, \textit{Physica B: Condensed Matter} {\bf{329}}, 1071-1073 (2003).

\bibitem{Fernandez} Teresa Medina Fernandez, \\ \textit{https://macsphere.mcmaster.ca/handle/11375/16485} (2014).

\bibitem{Abhishek1} Abhishek Juyal, Amit Agarwal, and Soumik Mukhopadhyay, \textit{Phys. Rev. Lett.} {\bf{120}}, 096801 (2018).

\bibitem{Abhishek2} Abhishek Juyal, Amit Agarwal, and Soumik Mukhopadhyay, \textit{Phys. Rev. B} {\bf{95}}, 125436 (2017).

\bibitem{Vinod7} Vinod Kumar Dwivedi, and Soumik Mukhopadhyay, \textit{AIP Conf. Proc.} {\bf{1832}}, 090016 (2017).

\bibitem{Vinod8} Vinod Kumar Dwivedi, and Soumik Mukhopadhyay, \textit{AIP Conf. Proc.} {\bf{1953}}, 120002 (2018).

\bibitem{Shannon} R. D. Shannon, \textit{Acta Cryst.} {\bf{A32}}, 751 (1976).

\bibitem{Rietveld1} H. M. Rietveld, \textit{J. AppL. Cryst.} {\bf{2}}, 65 (1969).

\bibitem{Aoyagi} Rintaro Aoyagi, Hiroaki Takeda, Soichiro Okamura and Tadashi Shiosaki, \textit{Jpn. J. Appl. Phys.} {\bf{40}}, 5671 (2001).

\bibitem{Yang1} W. C. Yang, Y. T. Xie, W. K. Zhu, K. Park, A. P. Chen, Y. Losovyj, Z. Li, H. M. Liu, M. Starr, J. A. Acosta, C. G. Tao, N. Li, Q. X. Jia, J. J. Heremans \& S. X. Zhang, \textit{Sci. Rep.} {\bf{7}}, 7740 (2017).

\bibitem{Gardner} Jason S. Gardner, Michel J. P. Gingras, and John E. Greedan, \textit{Rev. Mod. Phys.} {\bf{82}}, 53 (2010).

\bibitem{Bouchard} R. J. Bouchard and J. L. Gillson, \textit{Mater. Res. Bull.} {\bf{6}}, 669 (1971).

\bibitem{Baker} P. J. Baker, J. S. Moller, F. L. Pratt, W. Hayes, S. J. Blundell, T. Lancaster, T. F. Qi, and G. Cao, \textit{Phys. Rev. B} {\bf{87}}, 180409(R) (2013).

\bibitem{Qi} T F Qi, O B Korneta, Xiangang Wan, L E DeLong, P Schlottmann and G Cao, \textit{J. Phys.: Condens. Matter} {\bf{24}}, 345601 (2012).

\bibitem{Wan2} Xiangang Wan, Ashvin Vishwanath, and Sergey Y. Savrasov, \textit{Phys. Rev. Lett.} {\bf{108}}, 146601 (2012).

\bibitem{Blundell} Stephen Blundell, \textit{Magnetism in Condensed Matter, Oxford University Press, Great Britain}, 27-30 (2001).

\bibitem{Sirena} M. Sirena, L. B. Steren, and J. Guimpel, \textit{Phys. Rev. B} {\bf{64}}, 104409 (2001).

\bibitem{Mydosh} J. A. Mydosh, \textit{Rep. Prog. Phys.} {\bf{78}}, 052501 (2015).

\bibitem{Yoo} Y. J. Yoo, Y. P. Lee, J. S. Park, J.-H. Kang, J. Kim, B. W. Lee, and M. S. Seo, \textit{J. Appl. Phys.} {\bf{112}}, 013903(2012).

\bibitem{Kennedy} Brendan J. Kennedy, \textit{J. Solid State Chem.} {\bf{123}}, 14-20 (1996).

\bibitem{Singh} R. S. Singh, V. R. R. Medicherla, Kalobaran Maiti, and E. V. Sampathkumaran, \textit{Phys. Rev. B} {\bf{77}}, 201102(R) (2008).

\bibitem{Kallaher} R. L. Kallaher and J. J. Heremans, \textit{Phys. Rev. B} {\bf{79}}, 075322 (2009).

\bibitem{Yang2} W. C. Yang, W. K. Zhu, H. D. Zhou, L. Ling, E. S. Choi, M. Lee, Y. Losovyj, Chi-Ken Lu, and S. X. Zhang, \textit{Phys. Rev. B} {\bf{96}}, 094437 (2017).

\bibitem{Jafer} R. M. Jafer, E. Coetsee, A. Yousif, R. E. Kroon, O. M. Ntwaeaborwa, H. C. Swart, \textit{Appl. Surf. Sci} {\bf{332}}, 198-204 (2015).

\bibitem{Xinyu} Xinyu Zhang, Jiaqian Qin, Yanan Xue, Pengfei Yu, Bing Zhang, Limin Wang and Riping Liu, \textit{Sci. Rep.} {\bf{4}}, 4596 (2014). 

\bibitem{Harish3} Harish Kumar, V. G. Sathe, A. K. Pramanik, \textit{J. Magn. Magn. Mater} {\bf{478}}, 148-154 (2019).

\end{thebibliography}
\end{document}